\documentclass[conference]{IEEEtran}
\IEEEoverridecommandlockouts

\usepackage{cite}
\usepackage{amsmath,amssymb,amsfonts}
\usepackage{algorithmic}
\usepackage{graphicx}
\usepackage{textcomp}
\usepackage{xcolor}
\usepackage{multirow}
\usepackage{booktabs}
\usepackage{comment}
\usepackage{makecell}
\usepackage{threeparttable}
\usepackage{url}
\usepackage{tablefootnote}

\def\BibTeX{{\rm B\kern-.05em{\sc i\kern-.025em b}\kern-.08em
    T\kern-.1667em\lower.7ex\hbox{E}\kern-.125emX}}
\makeatletter
\newcommand{\linebreakand}{%
  \end{@IEEEauthorhalign}
  \hfill\mbox{}\par
  \mbox{}\hfill\begin{@IEEEauthorhalign}
}
\makeatother

\begin{document}

\title{Diarization-Aware Multi-Speaker Automatic Speech Recognition via Large Language Models\\
\thanks{Corresponding author: Ming Li, E-mail: ming.li369@dukekunshan.edu.cn}
}

\author{
\IEEEauthorblockN{
Yuke~Lin\IEEEauthorrefmark{1}\IEEEauthorrefmark{2}, 
Ming~Cheng\IEEEauthorrefmark{1}\IEEEauthorrefmark{2}, 
Ze~Li\IEEEauthorrefmark{1}\IEEEauthorrefmark{2},
Beilong~Tang\IEEEauthorrefmark{2}, 
Ming~Li\IEEEauthorrefmark{1}\IEEEauthorrefmark{2}, 
}
\IEEEauthorblockA{\IEEEauthorrefmark{1}School of Computer Science, Wuhan University, China}
\IEEEauthorblockA{\IEEEauthorrefmark{2}Suzhou Municipal Key Laboratory of  Multimodal Intelligent Systems, Digital Innovation Research Center,\\ Duke Kunshan University, China}
}

\maketitle

\begin{abstract}
Multi-speaker automatic speech recognition (MS-ASR) faces significant challenges in transcribing overlapped speech, a task critical for applications like meeting transcription and conversational analysis. While serialized output training (SOT)-style methods serve as common solutions, they often discard absolute timing information, limiting their utility in time-sensitive scenarios. Leveraging recent advances in large language models (LLMs) for conversational audio processing, we propose a novel diarization-aware multi-speaker ASR system that integrates speaker diarization with LLM-based transcription. Our framework processes structured diarization inputs alongside frame-level speaker and semantic embeddings, enabling the LLM to generate segment-level transcriptions. Experiments demonstrate that the system achieves robust performance in multilingual dyadic conversations and excels in complex, high-overlap multi-speaker meeting scenarios. This work highlights the potential of LLMs as unified back-ends for joint speaker-aware segmentation and transcription.

\end{abstract}


\section{Introduction}

Automatic Speech Recognition (ASR)~\cite{asr01, asr02, asr03} for single-speaker scenarios has achieved remarkable success and widespread deployment in industrial applications. However, real-world conversations, such as meetings, interviews, and discussions, often involve multiple speakers. In such settings, conventional ASR systems fail to distinguish who spoke what, rendering them insufficient for tasks requiring speaker-specific content understanding, such as meeting summarization and speaker-centric analysis. 

Multi-Speaker ASR (MS-ASR)~\cite{MS-ASR_survey, ms-asr-first-type01, ms-asr-first-type02, ms-asr-first-type03, ms-asr-second-type01, ms-asr-second-type02} extends the conventional ASR task by not only transcribing speech content but also attributing each utterance to the correct speaker. Unlike standard ASR, which assumes a single active speaker, MS-ASR must operate in conversational scenarios where multiple speakers take turns or speak simultaneously. The core challenge lies in performing accurate transcription under these complex interaction patterns, particularly when no prior knowledge about speakers' number, order, or identity is available. This requires the system to distinguish and organize utterances by the speaker while maintaining the transcription quality expected from modern ASR models.

\begin{figure*}[t]
  \includegraphics[width=0.95\textwidth]{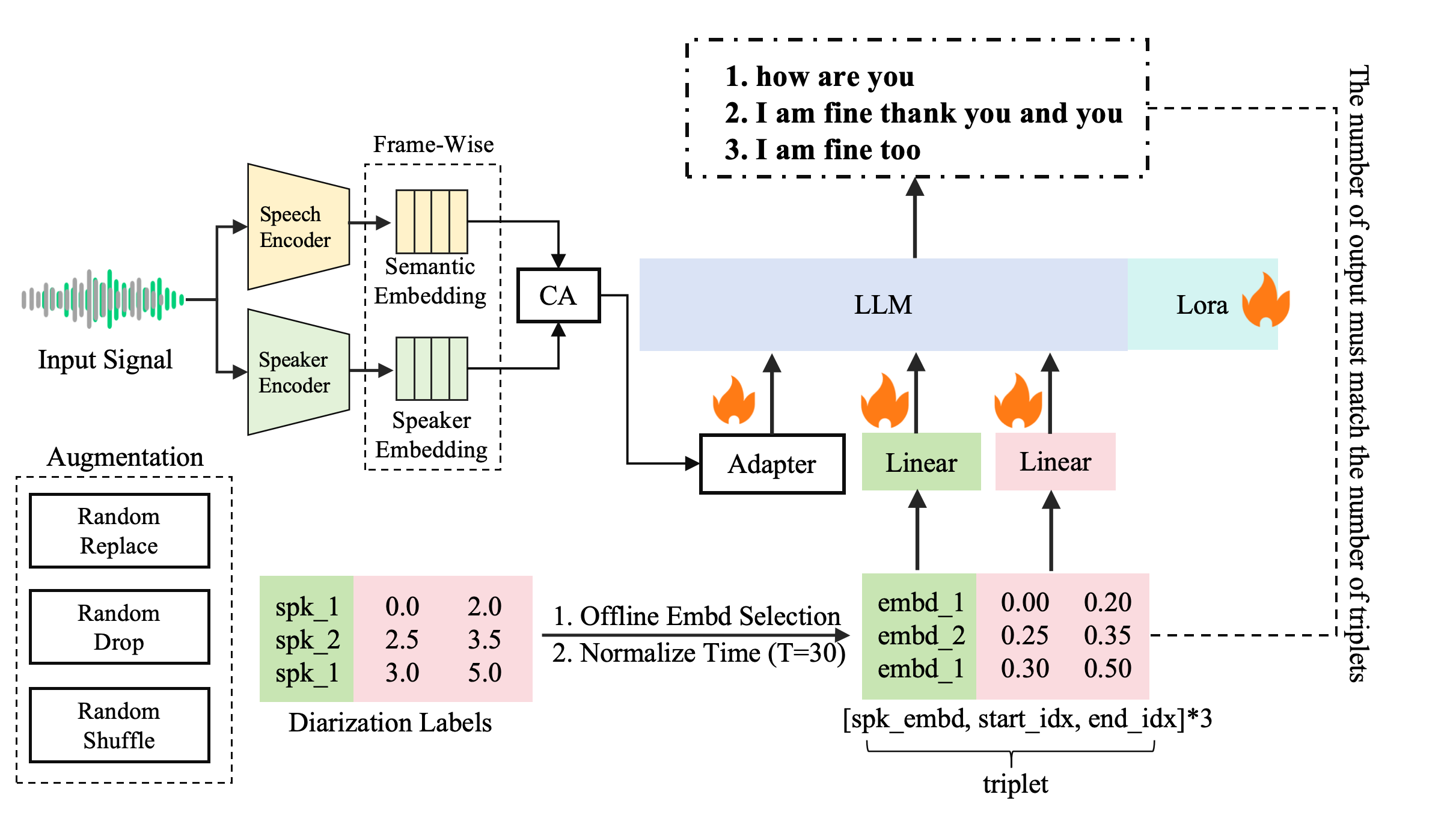}
  \centering
  \caption{{\it An overview of our framework. }}
  \label{fig::backbone}
  \vspace{-1.2em}
\end{figure*}

Early approaches to MS-ASR typically follow a modular pipeline design and can be broadly categorized into two types. The first type relies on speaker diarization to split the audio into multiple segments based on the predicted target-speaker voice activities~\cite{ms-asr-first-type01, ms-asr-first-type02, ms-asr-first-type03}. These segments are then fed into a standard ASR model for transcription. In overlapping regions, additional speech separation techniques~\cite{ss01, ss02, ss03} can be applied prior to transcription. The second type of pipeline performs speaker diarization and ASR independently and in parallel~\cite{ms-asr-second-type01, ms-asr-second-type02}. After obtaining transcriptions from the ASR system, time-aligned word boundaries are estimated through the forced alignment technique. These timestamped transcriptions are then matched to diarization outputs via a so-called orchestration process, which attempts to attribute each utterance to a speaker label based on temporal information. However, both paradigms leverage well-established components and suffer from several drawbacks. Errors from one module (e.g., diarization or alignment inaccuracies) can propagate to the final output. Also, the independent components are trained separately and lack a shared optimization objective.

More recent advances include end-to-end frameworks such as Permutation Invariant Training (PIT)~\cite{pit01, pit02, pit03} and Serialized Output Training (SOT)~\cite{sot01, sot02, sot03, t-sot}. However, due to the complexity of PIT, the performance of PIT-based systems tends to degrade as the number of speakers increases. In contrast, SOT-style models avoid limiting the maximum number of speakers. Nevertheless, since related content from different speakers is concatenated, and the grammatical structure of utterances in meeting scenarios is often suboptimal, these models require strong long-context awareness and cross-utterance modeling. Furthermore, they lack temporal alignment capabilities, which prevents them from producing time-stamped transcriptions. This type of work, which involves predicting only speaker labels and transcriptions without considering timing information, is usually referred to as Speaker-Attributed ASR (SA-ASR)~\cite{t-sot}.

Semi-End-to-End (Semi-E2E) approaches, such as Target-Speaker ASR (TS-ASR)~\cite{ts-asr01, ts-asr02, ts-asr03, ts-asr04, ts-asr05}, have emerged as a promising alternative. These models accept speaker embedding as input and generate transcriptions for that target speaker, allowing the prediction of the results for each speaker individually. However, most TS-ASR systems face two limitations: (1) the transcribed text cannot be temporally aligned with speaker diarization outputs; (2) they often operate on one speaker at a time, missing contextual information in conversations.

In this work, we propose a new semi-end-to-end MS-ASR framework that addresses the limitations of TS-ASR by introducing a triplet-based enrollment mechanism, where each enrolled target speaker is represented by a tuple consisting of (i) a speaker embedding, (ii) a start time of sentence, and (iii) an end time of sentence. This representation allows the model to jointly transcribe speech and output utterance-level timestamps that are directly aligned with a diarization system. Furthermore, we extend the MS-ASR architecture by integrating the large language models (LLMs)~\cite{llama, deepseek, qwen3}, enabling the system to simultaneously process multiple enrolled speaker triplets. This design facilitates contextual modeling across speakers and utterances, improving the efficiency and accuracy of transcription in highly overlapped multi-speaker scenarios.

We evaluate our system in the AliMeeting dataset~\cite{alimeeting}, a real-world Mandarin meeting corpus, and further test its generalization to the MLC-SLM Challenge~\footnote{\url{https://www.nexdata.ai/competition/mlc-slm}} - Task 2. We adopt tcpWER~\cite{tcpwer} as our primary evaluation metric, which jointly measures transcription accuracy and temporal alignment. Experimental results demonstrate the effectiveness of our method in producing high-quality, speaker-attributed transcriptions in challenging conversational settings.

\section{Method}
\subsection{Model Backbone}
As illustrated in Fig.~\ref{fig::backbone}, our framework integrates a large language model (LLM) with two parallel frame-wise encoders:
\begin{itemize}
    \item A speech encoder that extracts semantic embeddings $\mathbf{H}^s = [\mathbf{h}_1^s, \mathbf{h}_2^s, ..., \mathbf{h}_T^s] \in \mathbb{R}^{T \times d_s}$ from the input audio, where $T$ denotes the number of frames and $d_s$ the semantic embedding dimension.
    \item A speaker encoder that produces speaker-discriminative features $\mathbf{H}^p = [\mathbf{h}_1^p, \mathbf{h}_2^p, ..., \mathbf{h}_T^p] \in \mathbb{R}^{T \times d_p}$, where $d_p$ denotes the speaker embedding dimension.
\end{itemize}

To enable speaker-adaptive semantic modeling, we propose a gated cross-attention mechanism that dynamically integrates speaker information with semantic features. The process consists of two key steps: First, semantic features $\mathbf{H}^s$ (from the speech encoder) attend to speaker features $\mathbf{H}^p$ through cross-attention:
\begin{equation}\label{eq::cross-attention}
\mathbf{H}^{ca} = \text{Cross-Attention}(\mathbf{Q}=\mathbf{H}^s, \mathbf{K}=\mathbf{H}^p, \mathbf{V}=\mathbf{H}^p)
\end{equation}

\noindent where $\mathbf{H}^s$ serves as queries to selectively aggregate relevant speaker characteristics from $\mathbf{H}^p$. Second, the attended features are then adaptively gated and combined with the original semantic features:
\begin{equation}
\mathbf{H}^{o} = \sigma(\mathbf{W}_g\mathbf{H}^{ca}) \odot \mathbf{H}^{ca} + \mathbf{H}^s
\end{equation}

This gating mechanism, controlled by the sigmoid function $\sigma \left ( \cdot  \right )$, allows flexible modulation of speaker influences while the residual connection preserves essential semantic information. Finally, refined features $\mathbf{H}^{o}$ are projected through adapter layers to align with the input space of the LLM.

\subsection{Diarization-Aware Triplet Enrollment}

\begin{figure}[t]
  \includegraphics[width=0.45\textwidth]{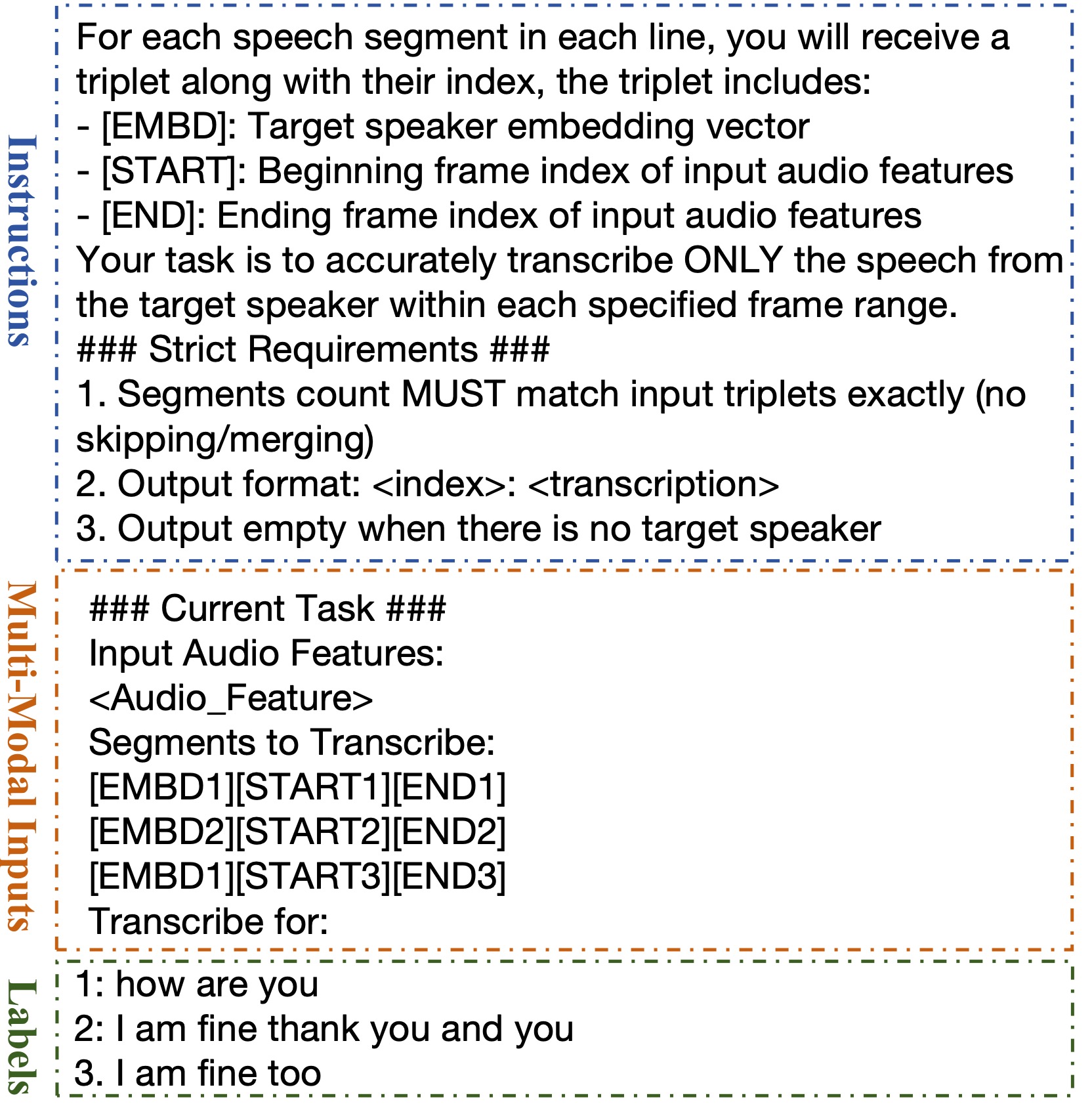}
  \centering
  \caption{{\it The construction of our inputs (three sentences from two speakers).}}
  \label{fig::instruction}
  \vspace{-1.2em}
\end{figure}

As shown in Fig.~\ref{fig::instruction}, the whole inputs can be roughly divided into three parts: instructions, multi-modal inputs and labels. The instructions provide the objective and constraints of our task, and the labels are used for generation. As for the multi-modal inputs, to explicitly incorporate diarization awareness into the LLM, we construct a structured triplet representation comprising (1) the target speaker embedding obtained through offline embedding selection and (2) normalized start and end times computed as $\text{frame index} / \text{total frames}$ for the audio chunk, where both temporal boundaries undergo identical normalization. As illustrated in Fig.~\ref{fig::backbone}, this triplet formulation captures essential diarization elements - speaker identity and precise temporal boundaries - in a unified representation that facilitates LLM integration.

These triplets describe speaker identity and utterance-level time boundaries are linearly projected into the LLM input space. The triplet inputs are fed as conditioning instructions, guiding the model in decoding the speech content corresponding to the given speaker and time interval. Unlike traditional TS-ASR methods that rely solely on speaker embeddings or time ranges, our approach combines both, enabling finer speaker-utterance disambiguation, especially under heavily overlapped conditions. Moreover, since the LLM can process multiple triplets simultaneously, our system supports joint decoding of multiple utterances from multiple speakers, with contextual modeling across speaker turns.

\subsection{Data Augmentation Strategies}

To improve robustness against diarization errors, we propose three data augmentation techniques during training.

\subsubsection{Embedding Replacement}
During training, we randomly replace the original speaker embeddings in some triplets with other speaker embeddings (probability: $p_{replace}$), while setting their corresponding transcription labels to empty. This simulates the imperfect speaker embedding extraction in real-world scenarios and forces the model to ignore incorrect speaker information.

\subsubsection{Embedding Dropout}
During training, we randomly drop entire triplets (probability: $p_{drop}$) and their associated labels. This simulates the presence of missing diarization outputs and forces the model to strictly follow the available diarization input rather than attempting to transcribe all speech content, ensuring output alignment with given triplet sequences.

\subsubsection{Triplet Shuffling}
During training, we randomly shuffle the input triplets (probability: $p_{shuffle}$) while simultaneously reordering their corresponding labels. This guarantees that the model aligns outputs with the given enrollment order, regardless of actual chronological order.

\subsection{Chunk-Based Inference}
To handle long-form recordings, we implement chunk-based inference. The system first splits prolonged speaker segments into fixed-duration chunks (default: 30s), then organizes them into coherent processing units while enforcing constraints on maximum chunk duration, speaker segment counts per chunk, and total segments per chunk. Cross-chunk continuity is preserved by maintaining consistent speaker embeddings and temporal alignment. This design ensures scalable processing of extended conversations while retaining accurate diarization-to-transcription mapping.

\section{Experimental Settings}
\subsection{Dataset Usage}

To support the pretraining of large-scale neural networks for multi-speaker ASR, we construct a synthetic dataset using two publicly available corpora: Common Voice~\cite{common-voice} and VoxBlink2~\cite{lin2024voxblink,voxblink2}. Common Voice is a crowd-sourced multilingual ASR dataset; we use only its English subset, which provides high-quality transcribed utterances suitable for supervised ASR training. VoxBlink2 is a large-scale speaker recognition dataset collected from YouTube. Although it lacks transcriptions, it offers substantial speaker diversity and is suitable for generating speaker embeddings. To enable supervised learning, all VoxBlink2 utterances are transcribed using Whisper-Large-V3 model~\cite{whisper-large-v3} as the ground truth, which finally obtains approximately 4 million high-quality utterance-text pairs, covering over 40,000 unique speakers. During training, we simulate multi-speaker conversational segments by randomly sampling and mixing utterances from different speakers. This mixing is performed on-the-fly to control the number of speakers within each mixture. This simulation strategy allows the model to pretrain on rich and diverse multi-speaker inputs with explicit speaker attribution, providing a strong foundation before fine-tuning on real-world conversational datasets.

To fine-tune our model on real-world conversational data, we adopt the AliMeeting dataset
, a publicly available Mandarin meeting corpus for highly-overlapped multi-talker scenarios. The corpus contains approximately 120 hours of recorded meetings involving 2 to 4 speakers per session, covering diverse room sizes and acoustic conditions. We use only the single-channel recordings from the far-field 8-channel microphone array to match our intended deployment setting. Each session includes transcriptions with speaker and timing annotations. The training set includes 212 sessions ($\sim$105 hours), and the evaluation set includes 8 sessions ($\sim$4 hours), with average speaker overlap ratios of 42.3\% and 34.2\%, respectively. Lastly, an additional test set contains 20 sessions ($\sim$10 hours). All speakers are native Mandarin speakers engaged in natural discussions on various topics such as business, education, and healthcare. This dataset offers a realistic and challenging benchmark for multi-speaker speech recognition in Mandarin meeting scenarios, with rich acoustic variability, high overlap rates, and dialogue structures.

In addition to AliMeeting, we further fine-tune and evaluate our method on the dataset provided by the Interspeech 2025 Multilingual Conversational Speech Language Model Challenge (MLC-SLM). This dataset consists of two-speaker real-world conversations across 11 languages: English, French, German, Spanish, Japanese, Korean, and others. The conversations cover diverse topics and are recorded in quiet indoor environments using everyday devices such as smartphones. We focus on Task 2 of the challenge, which requires joint speaker diarization and ASR without oracle segmentation or speaker labels. This makes the task significantly more challenging, as the system must autonomously determine "who spoke what and when" from unsegmented audio. Compared with the highly overlapped speech and multiple speakers in the AliMeeting dataset, the MLC-SLM dataset features low-overlap, two-speaker dialogues but introduces significant linguistic diversity. By leveraging this dataset, we aim to demonstrate the capability of our approach to generalize to multilingual scenarios.

\subsection{Diarization System}

To obtain speaker embeddings and utterance-level time boundaries required by our MS-ASR framework, we adopt the Sequence-to-Sequence Neural Diarization (S2SND)~\cite{s2snd} system as the prior speaker diarization module. S2SND builds upon the authors’ prior work on Sequence-to-Sequence Target-Speaker VAD (Seq2Seq-TSVAD)~\cite{s2s-tsvad}, and extends it to a more general diarization framework that supports both online and offline inference without relying on clustering or permutation-invariant training. In our work, we use the S2SND-Small variant containing 16.56 million parameters. Following the original training strategy described in the paper, we train this model on the AliMeeting and MLC-SLM datasets. The generated offline diarization outputs are used to extract (i) per-speaker embeddings and (ii) sentence-level start/end timestamps. These outputs are further used as enrollment triplets in our MS-ASR system.

\subsection{Multi-Speaker ASR Training}
\subsubsection{Model Structure}

Our experimental setup employs Qwen2.5-3B~\cite{qwen2.5} as the foundation model for fine-tuning, enhanced with a multi-modal architecture for speech processing. The audio pipeline utilizes the frozen encoder from Whisper-large-v3-turbo~\cite{whisper-large-v3} as our primary speech feature extractor. At the same time, speaker representation is handled by a ResNet34~\cite{resnet34} network that generates both frame-level embeddings and utterance-level embeddings for offline scenarios. The multi-modal adapter integrates a hierarchical structure consisting of three 1D convolutional layers (with kernel size 3 and progressively increasing strides of 1, 2, and 2) followed by three transformer layers featuring a 640-dimensional attention mechanism, culminating in a linear projection layer that aligns the feature space with the LLM's input dimensions. For efficient parameter adaptation, we implement Low-Rank Adaptation (LoRA) with a rank of 8 and alpha of 16, maintaining the frozen state of all speech components throughout training to ensure stable acoustic feature extraction while enabling effective language model adaptation.

\subsubsection{Training Data}
The training data consists of two complementary sources: simulated and real conversational data. For simulated data, we generate multi-speaker mixtures by randomly combining utterances from different speakers, inserting natural silence intervals, and controlling overlap patterns. The real data pipeline extracts continuous conversational segments while preserving authentic speaker dynamics and overlap structures. Both methods employ probabilistic augmentation through embedding replacement ($p_{replace}=0.05$), embedding dropout ($p_{drop}=0.1$), and triplet shuffling ($p_{shuffle}=0.2$). We implement controls including maximum segment duration (30s), per-speaker utterance limits, and variable-length sampling windows. The utterance-wised speaker embeddings are randomly selected during training and mean-pooled during inference.

\subsubsection{Configuration}
Following the findings of~\cite{geng2025osum}, we implement a carefully designed two-stage training strategy to optimize resource efficiency. The initial phase focuses exclusively on training the multi-modal adapter and associated triplet linear layers, while the subsequent stage introduces LoRA weights for supervised fine-tuning. Our hardware configuration utilizes 8 NVIDIA RTX A6000 GPUs with a per-device batch size of 2 and gradient accumulation steps of 4, effectively creating a larger aggregated batch size of 64. Given the limited number of training epochs, overfitting concerns are mitigated without requiring audio data augmentation. For evaluation, we employ distinct chunking strategies: the Alimeeting benchmark uses a segment limit of 10 with 4 segments per speaker, while the MLC-SLM evaluation adopts a configuration of 8 total segments with 6 segments per speaker, ensuring optimal performance for each specific task.

\subsection{Evaluation Metric}

For the speaker diarization, we report the Diarization Error Rate (DER). DER is computed as the sum of speaker confusion, missed speech, and false alarms divided by the total reference time, reflecting the diarization system's ability to accurately detect ``who spoke when."

For the multi-speaker ASR, we first use constrained permutation Word Error Rate (cpWER) as the metric. cpWER accounts for the permutation ambiguity between speaker labels during evaluation by considering all possible alignments between system hypotheses and reference transcriptions. It is commonly used in settings where reference speaker identities are known, but the system may produce transcriptions in an arbitrary speaker order.

To further evaluate the temporal accuracy of our system, we also report time-constrained permutation WER (tcpWER)~\cite{tcpwer}. This metric extends cpWER by incorporating temporal alignment constraints between predicted and reference utterances, ensuring that the content and timing are jointly correct. Specifically, tcpWER penalizes misaligned transcriptions even if the word sequence is correct, making it more suitable for speaker-attributed ASR tasks where timing precision is essential.

\begin{table}[t]
    \setlength{\tabcolsep}{8pt}
    \renewcommand{\arraystretch}{1.2}
    \centering
    \caption{cpWER (\%) of our proposed system and existing methods on the AliMeeting Eval and Test sets.}
    \label{tab:alimeeting_cpwer}
    \begin{tabular}{lrr}
    \toprule
    \textbf{Method} & \textbf{Eval} & \textbf{Test} \\
    \midrule
    Cascaded SA-ASR &  ~ & ~ \\
    \quad FD-SOT~\cite{yu22b_interspeech} & 41.0 &41.2 \\
    \quad WD-SOT~\cite{yu22b_interspeech} & 36.0 &37.1 \\
    CASA-ASR~\cite{shi23d_interspeech} & 31.8 & 34.7 \\
    SA-Paraformer~\cite{10389762} & 36.2 & 38.6 \\
    \quad+interCTC with f\&i-speaker &  32.5 & 34.8 \\
    \midrule
    \textbf{Ours} & 31.6 & 35.1 \\
    \bottomrule
\end{tabular}
\end{table}

\section{Results}

To support our MS-ASR system, we first evaluate the performance of the prior speaker diarization module. Following the official evaluation protocol of the AliMeeting benchmark, which adopts Oracle voice activity detection (VAD) and a 0.25-second collar tolerance, our S2SND-based diarization system achieves a DER of 4.96\% on the Eval set and 3.77\% on the Test set. These outputs are used in all subsequent experiments on the AliMeeting dataset. For the MLC-SLM Challenge, which does not provide Oracle VAD and disallows any collar tolerance, our diarization module yields a DER of 14.27\% on the Dev set. The ground truth annotations for the Test set remain unavailable, so no direct DER evaluation is accessible. These outputs are used for all experiments involving the MLC-SLM dataset.

Table~\ref{tab:alimeeting_cpwer} presents our proposed system's cpWER (\%) performance compared with several state-of-the-art multi-speaker ASR baselines on the AliMeeting Eval and Test sets. While these referenced works use the speaker-dependent Character Error Rate (SD-CER) to emphasize evaluation at the recording level with consistent speaker label permutation, this setting is now aligned with the current cpWER protocol. Therefore, we treat cpWER and SD-CER as equivalent in this context. Our system achieves the lowest cpWER on the Eval set (31.6\%), outperforming the strong CASA-ASR~\cite{shi23d_interspeech} and interCTC-based Paraformer variants~\cite{10389762}. On the Test set, our system achieves 35.1\%, slightly higher than CASA-ASR (34.7\%) but still competitive.

It is worth noting that unlike the other existing systems, which are optimized solely for speech recognition accuracy, our model is designed to generate speaker-attributed transcriptions with accurate utterance-level timestamps jointly. Therefore, a marginal difference in cpWER is acceptable, given our system's richer and more structured output. These results validate the effectiveness of our diarization-aware MS-ASR framework. Specifically, the use of designed triplet enrollment enables precise utterance segmentation per speaker, while the LLM-based decoder allows joint decoding of multiple speakers with contextual modeling. This architecture is well-suited for real-world meeting transcription scenarios requiring transcript accuracy and timing alignment.

\begin{table}[t]
    \setlength{\tabcolsep}{20pt}
    \renewcommand{\arraystretch}{1.2}
    \centering
    \caption{tcpWER (\%) of our proposed system on the AliMeeting Eval and Test sets by number of speakers.}
    \label{tab:alimeeting_tcpwer}
    \begin{tabular}{lrr}
    \toprule
    \textbf{Num. Spks} & \textbf{Eval} & \textbf{Test} \\
    \midrule
    2  & 14.94 & 13.57 \\
    3  & 31.58 & 29.73 \\
    4  & 41.59 & 52.67 \\
    \midrule
    Overall & 32.17 & 36.36 \\
    \bottomrule
\end{tabular}
\end{table}

Table~\ref{tab:alimeeting_tcpwer} reports the performance of our system on the AliMeeting Eval and Test sets using the time-constrained permutation WER (tcpWER), which simultaneously evaluates transcription accuracy and temporal alignment. Under the evaluation metric setting, a prediction is penalized even if its word content is correct as long as the predicted utterance time deviates from the reference by more than 5 seconds. This makes tcpWER a stricter and more comprehensive metric than conventional WER or cpWER. To our knowledge, our work is the first to report tcpWER results on the AliMeeting dataset. As shown in the table, the overall tcpWER is 32.17\% on the Eval set and 36.36\% on the Test set, reflecting our system’s ability to produce not only accurate transcriptions but also reliable utterance-level timestamps.

We further break down the results by the number of speakers in each recording. As expected, the tcpWER increases with speaker count: from 14.94\% / 13.57\% (2 speakers) to 31.58\% / 29.73\% (3 speakers), and further to 41.59\% / 52.67\% (4 speakers) on the Eval/Test sets. This trend reflects the increased challenge of speaker-attributed transcription under higher overlap and more frequent speaker turns. Nonetheless, our proposed system maintains reasonable performance even in the most challenging 4-speaker scenarios, demonstrating its robustness to multi-party scenarios. These results validate our design choice of combining speaker embeddings with speaking-time information to anchor utterances precisely. The tcpWER evaluation further highlights our system’s capability to meet real-world demands for timestamped and speaker-attributed transcription in long-form multi-speaker conversations.

Table~\ref{tab:mlc_slm} presents the tcpWER (\%) of our proposed system compared to the official baseline on the dataset of MLC-SLM Challenge - Task 2, evaluated on both the Dev and Test sets. As the challenge requires, tcpWER is also computed with a 5-second tolerance window, jointly evaluating both transcription accuracy and temporal alignment. On the Dev set, our system achieves substantial improvements over the baseline across all 15 evaluated languages. The average tcpWER is reduced from 76.12\% to 24.95\%, demonstrating the effectiveness of our diarization-aware MS-ASR pipeline. The improvement is particularly notable for low-resource and morphologically rich languages such as Portuguese (from 118.84\% to 37.35\%), French (from 96.04\% to 34.74\%), and German (from 86.74\% to 30.38\%). These results suggest that our model can generalize to diverse language patterns, even under challenging multilingual scenarios.

Moreover, the English language is evaluated across five regional variants according to the challenge protocol. Our model consistently lowers tcpWER across all English subsets, achieving reductions of more than 50\% absolute in most cases, with the lowest error observed for English-Australian (14.40\%) and English-Indian (16.25\%). On the Test set, where only the average tcpWER is available through the official evaluation server, our model also significantly outperforms the baseline, reducing the score from 60.39\% to 20.44\%. This result further confirms the robustness and transferability of our approach in real-world multilingual diarization and ASR tasks.

\section{Conclusions}

This work presents a novel semi-end-to-end framework for multi-speaker ASR that integrates diarization-aware inputs with large language models (LLMs). Our method introduces a triplet-enrollment instruction design, combining speaker embeddings and utterance-level timing information to guide the transcription process. Unlike previous TS-ASR approaches that rely solely on speaker identity or time-range cues, our framework enables precise disambiguation of overlapping utterances, supporting the joint decoding of multiple speakers with accurate temporal alignment.

Extensive experiments on the AliMeeting dataset show that our system achieves state-of-the-art cpWER on the evaluation set and delivers competitive performance on the test set. More significantly, it produces high-quality utterance-level timestamps, as evidenced by strong tcpWER results — a metric we report for the first time on this benchmark. Additionally, experiments on the MLC-SLM Challenge demonstrate the broad generalizability of our method, with substantial tcpWER reductions across a wide range of languages.

In future work, we plan to investigate the streaming inference strategies for real-time deployment and expand the system's applicability to downstream tasks such as meeting summarization and speaker-centric question answering.

\begin{table}[t]
    \setlength{\tabcolsep}{6pt}
    \renewcommand{\arraystretch}{1.2}
    \centering
    \caption{Performance Comparison (tcpWER \%) on the MLC-SLM Dev and Test sets by different languages.}
    \label{tab:mlc_slm}
    \begin{tabular}{lcccc}
    \toprule
    \multirow{2}{*}{\textbf{Language}} & \multicolumn{2}{c}{\textbf{Official Baseline~\tablefootnote{\url{https://github.com/mubingshen/MLC-SLM-Baseline/tree/main}}}} & \multicolumn{2}{c}{\textbf{Ours}}
    \\
    \cmidrule(lr){2-3}
    \cmidrule(lr){4-5}
    & \textbf{Dev} & \textbf{Test} & \textbf{Dev} & \textbf{Test}
    \\
    \midrule
    English-American & 53.73 & - & 23.01 & - \\
    English-Australian & 52.63 & - & 14.40 & - \\
    English-British & 71.92 & - & 18.69 & - \\
    English-Filipino   & 50.37 & -  & 18.14 & - \\
    English-Indian   & 70.72 & -  & 16.25 & - \\
    French   & 96.04 & -  & 34.74 & - \\
    German   & 86.74 & -  & 30.38 & - \\
    Italian   & 83.31 & -  & 19.90 & - \\
    Japanese   & 71.30 & -  & 36.29 & - \\
    Korean   & 59.55 & -  & 27.04 & - \\
    Portuguese   & 118.84 & -  & 37.35 & - \\
    Russian   & 69.21 & -  & 23.20 & - \\
    Spanish   & 75.61 & -  & 23.17 & - \\
    Thai   & 83.56 & -  & 20.93 & - \\
    Vietnamese   & 82.80 & -  & 29.76 & - \\
    \midrule
    Overall & 76.12 & 60.39 & 24.95 & 20.44 \\
    \bottomrule
\end{tabular}
\end{table}

\section*{Acknowledgments}
This research is funded in part by the National Natural Science Foundation of China (62171207), Yangtze River Delta Science and Technology Innovation Community Joint Research Project (2024CSJGG01100). Many thanks for the computational resource provided by the Advanced Computing East China Sub-Center.

\bibliographystyle{IEEEtran}
\bibliography{ref}

\end{document}